# Enhancing Multiagent Genetic Network Programming Performance Using Search Space Reduction

**Ali Kohan [1]  Mohamad Roshanzamir [1]  Roohallah Alizadehsani [2]**

[1] Department of Computer Engineering, Faculty of Engineering, Fasa University, Fasa, Iran

a-kohan971@fasau.ac.ir
roshanzamir@fasau.ac.ir

[2] Institute for Intelligent Systems Research and Innovation (IISRI), Deakin University, Geelong, Australia

r.alizadehsani@deakin.edu.au

## Abstract

Genetic Network Programming (GNP) is an evolutionary algorithm that extends Genetic Programming (GP). It is typically used in agent control problems. In contrast to GP, which employs a tree structure, GNP utilizes a directed graph structure. During the evolutionary process, the connections between nodes change to discover the optimal strategy. Due to the large number of node connections, GNP has a large search space, making it challenging to identify an appropriate graph structure. One way to reduce this search space is by utilizing simplified operators that restrict the changeable node connections to those participating in the fitness function. However, this method has not been applied to GNP structures that use separate graphs for each agent, such as situation-based GNP (SBGNP). This paper proposes a method to apply simplified operators to SBGNP. To evaluate the performance of this method, we tested it on the Tileworld benchmark, where the algorithm demonstrated improvements in average fitness.

KEYWORDS: Evolutionary Algorithms, Agent Control Problems, Genetic Programming, Genetic Network Programming (GNP), Search Space Reduction, Explainable Artificial Intelligence (Explainable AI)

## 1. Introduction

Genetic Network Programming (GNP) [1] is an evolutionary computation technique that builds on the foundations of Genetic Programming (GP). It utilizes a network structure, specifically a directed graph, instead of the bit-string structure used in Genetic Algorithms (GA) or the tree structure used in Genetic Programming (GP). This network-based approach allows for more efficient handling of complex problems by enabling the reuse of nodes and providing flexible transitions between states. In the optimization context, the No-Free-Lunch (NFL) theorem [2] states that no single algorithm can outperform all others across every possible problem space. This principle emphasizes the importance of selecting the appropriate algorithm based on the problem. Therefore, employing both tree-based and graph-based representations can enhance problem-solving capabilities by leveraging the strengths of each structure for different types of problems.

GNP has demonstrated its capability in various domains, such as stock trading [3-6], robot control [7-9], social networks [10], etc. A key advantage of GNP is its interpretability, especially when compared to deep learning models and most machine learning models used nowadays [11]. This focus on improving the explainability of models in artificial intelligence (AI) is known as explainable AI (XAI). This has become increasingly critical in recent years, particularly in fields such as healthcare [12], autonomous vehicles [13], and finance [14].

In general, GNP uses three types of nodes. The interplay between these nodes enables the GNP framework to model decision-making processes in complex problems. These node types are as follows:

- The **Start node** determines the initial point of execution in the program, serving as the entry point for running the network.

- **Judgment nodes** operate as conditional decision-making points, similar to if-then statements. These nodes analyze the environment and determine the current state, directing the program flow by selecting appropriate actions based on the conditions they assess.
- **Processing nodes** execute the actions. Unlike judgment nodes, processing nodes do not involve conditional branching; they perform predetermined operations without evaluating conditions.

The GNP structure can be demonstrated in two forms: genotype and phenotype. As illustrated in Figure 1, the phenotype form represents a directed graph. The agent uses one of these nodes at each step, transitioning between them until the program terminates. In genotype form, each node is represented by its features in bit-string format. For node $i$: $NT_i$ denotes the node type, with values of 0, 1, and 2 corresponding to start, judgment, and processing nodes, respectively. $NF_i$ indicates the specific function to be executed. $d_i$ represents the amount of time required to perform this function. $C_{ij}$ denotes the node to which the jth branch of node i is connected. $d_{ij}$ represents the time needed to move from the current node to node $C_{ij}$.

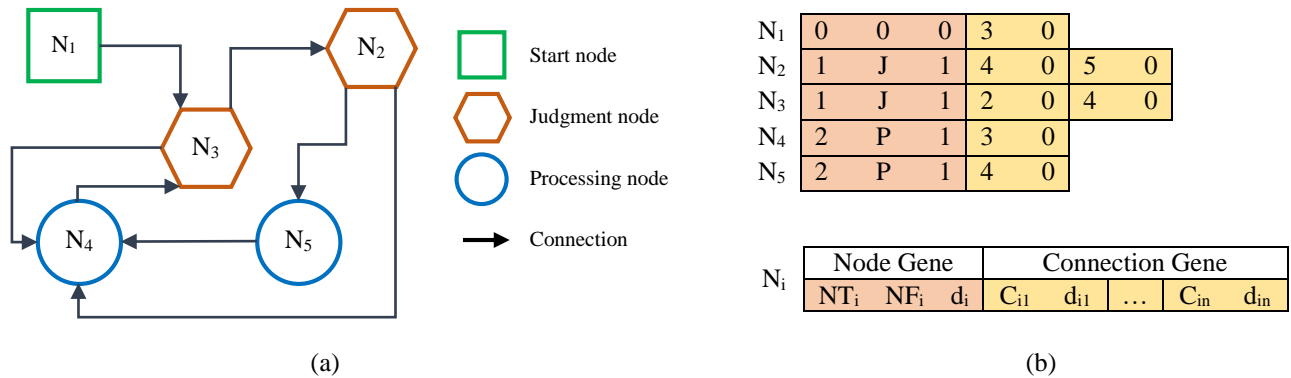

Figure 1. An example of a) phenotype and b) genotype representation in GNP.

This paper is organized as follows. The following section briefly describes some of the most commonly used GNP variants. Next, in Section 3, the proposed algorithm is presented. Then, experimental results and the proposed method's performance compared to GNP, GNP_simplified, and SB-GNP are presented in Section 4. Finally, the conclusion and future work are discussed in Section 5.

## 2. Related works

Much research has focused on optimizing the GNP operation (like crossover and mutation) and combining it with other machine learning methods like Reinforcement Learning (RL) to enhance the efficiency and adaptability of GNP. Much research has been conducted on using RL to enable systems to learn and evolve more effectively from their interactions with the environment [9, 15-17]. Mabu et al. [18] introduced the first modification to GNP by incorporating online learning through Q-learning [19] to enhance the transfer rules between states. This approach was tested on the prisoner's dilemma problem. Similar modifications were made in [20, 21], where Q-learning was combined with GNP to improve its adaptability in dynamic environments. Alternatively, some studies [22, 23] employed the SARSA algorithm [19] instead of Q-learning and applied it to optimize the control process of Khepera robots [24]. In another study, a new algorithm called Probabilistic Model Building Genetic Network Programming (PMBGNP) was introduced by Li et al. [17]. This algorithm utilizes the graph structure of GNP to provide superior solution representation compared to traditional estimation of distribution algorithms for specific problems. Furthermore, a variant of this algorithm, known as reinforced PMBGNP, has been developed to combine PMBGNP with reinforcement learning, aiming to improve performance regarding fitness values, search speed, and reliability. These algorithms are used to address the challenges of controlling agents' behavior.

Another area of research aimed at enhancing the GNP mechanism is the combination of swarm intelligence and GNP. Due to the random nature of Genetic Algorithms (GA) and their inherent exploration characteristics in the environment, an approach to preserve beneficial genes in individuals can potentially improve the evolutionary process. In [25, 26], the Ant Colony Optimization (ACO) [27] was combined with GNP to enhance the exploitation ability of GNP. In general, GNP biases to exploration and ACO biases to exploitation [28]. One iteration of ACO is executed after every ten iterations of

GNP to integrate the strengths of both GNP and ACO algorithms. However, since crossover and mutations remained unchanged, the destruction of valuable structures continued to occur. Another swarm intelligence algorithm integrated with GNP is Artificial Bee Colony (ABC) [29]. In [30], an ABC-based evolution strategy was used. The algorithm consisted of three phases: Employed bee phase, Onlooker bee phase, and Scout bee phase. During the Employed and Onlooker bee phases, individuals in the population engage in knowledge sharing, creating new individuals. In contrast, during the Scout bee phase, individuals that fail to demonstrate improvement after a predetermined number of iterations are replaced by new individuals. The newly generated individual will survive if it outperforms the one that produced it.

The evolutionary process in GNP is guided by the frequent reuse of important nodes, a concept known as node reusability. Li et al. [31] explored this concept by analyzing node branches' reusability, treating it as a critical metric for evaluating node sets. The findings suggested that evolution pressure can be adjusted by modifying crossover and mutation probabilities. Specifically, the research showed that frequently used nodes and branches in high-fitness individuals should remain relatively stable, while low-fitness individuals should undergo more significant changes. This approach, called Adaptive GNP, introduced adaptive probabilities for crossover and mutation operations.

Two recently proposed algorithms that inspired this research are situation-based GNP (SBGNP) [32] and GNP_simplified [33]. They address two different drawbacks of standard GNP. GNP_simplified proposed considering the feature of "transition by necessity" and implementing a method to handle it. It uses simplified GNP operators instead of uniform ones used in standard GNP. In the crossover, each branch of the graph can contribute to evolution only if it is transited in at least one of the parent individuals. Similarly, each branch can contribute only in the mutation phase if it is transited during runtime. This approach reduces search space and computation time. As shown in Figure 2, the search space of GNP_simplified for crossover and mutation is more restricted compared to GNP. The orange area represents the search space in GNP_simplified, while the entire region (blue and orange) represents the search space in the standard GNP. *TB* and *UB* denote transited and untransited branches, respectively.

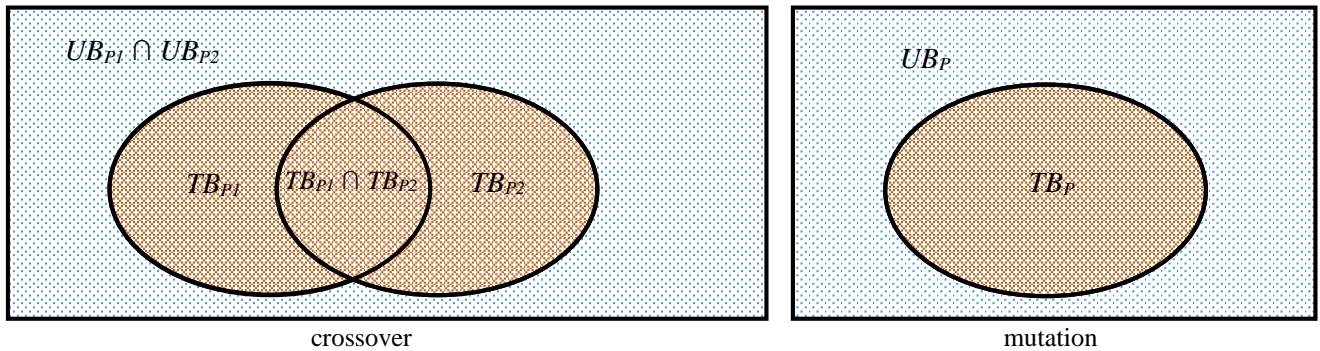

Figure 2. Search space in GNP_simplified operators.

On the other hand, SBGNP uses a novel approach where strategies for agents are generated individually based on their specific situations rather than seeking a single, optimal strategy for all agents. This is illustrated in Figure 3. This method simplifies the problem by aligning strategies to each agent's unique context, making it more efficient in achieving the desired goals. Finding a universal strategy for all agents is more complex than generating multiple strategies suited to individual circumstances. By employing this approach, the GNP algorithm gains greater flexibility in discovering better solutions.

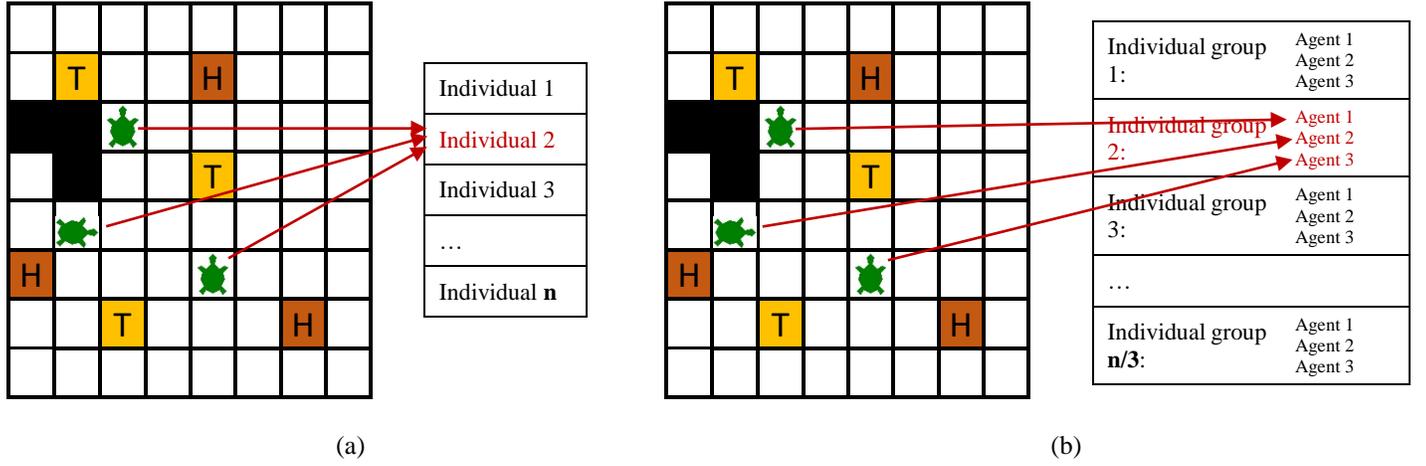

Figure 3. Comparison between individual structure in a) GNP and b) SBGNP.

## 3. Proposed algorithm

Inherently, GNP has a large search space, challenging its evolutionary process. Another issue with GNP is that all agents share the same graph, making it difficult to handle complex environments. However, using separate graphs for each agent would increase the search space even further, adding complexity. In GNP, the number of judgment and processing nodes is fixed and determined by the system designer. Throughout the evolution process, the node functions and the number of branches remain constant; only the node connections can be modified. Additionally, the number of instances for each node, referred to as the program size, is manually set and is typically the same for all nodes. For example, if the program size is set to five, it indicates that there are five instances of each judgment and processing node in every individual. The search space of GNP can be computed using Equation (1):

$$Search\_Space_{GNP} = (PS \times (NJ \times NB + NP))^{PS \times (NJ+NP)} \quad (1)$$

where the number of judgment functions is *NJ*, the number of processing functions is *NP*, and the program size is *PS*, with an average of *NB* branches per judgment function. This indicates that the search space of GNP is significantly large and complex. This complexity increases even further for algorithms that use separate graph structures for each agent, such as SBGNP [32]. For SBGNP, the search space is calculated by multiplying the number of agents by the search space obtained using Equation (1), as shown in Equation (2).

$$Search\_Space_{SBGNP} = Search\_Space_{GNP} \times NA \quad (2)$$

where *NA* is the number of agents.

The directed graph's node transitions enable the formulation of compact programs that solve problems. Specifically, these transitions facilitate the generation of sequential decision-making rules crucial for problem-solving. These rules follow the 'IF-THEN' structure, where only necessary judgment functions are executed to determine actions. The necessity of these judgments is determined through the evolutionary process. This 'transition by necessity' feature is a critical characteristic that is not considered in standard GNP. It can result in invalid/negative evolution. Considering this concept, we utilize simplified operators [33] to restrict the search space and enable more efficient evolution by reducing invalid/negative evolution in the GNPs using separate graphs for each agent. Simplified operators reduce the search space by limiting the possible changes in branches to only the transited branches.

The algorithm of the proposed method is shown in Algorithm 1. Firstly, because we want to consider separate graph structures for each agent, we need to group the initial population into *n* groups. Each group contains the individuals (i.e., graph structures) related to each agent. For example, if we have three agents, each group has three individuals. The main loop begins by running each group in the environment. The agents take turns taking one step, one by one and then repeat the process. The transited branches are stored for the evolution operations. Then, according to the fitness values, two parents are selected for the crossover operation. In this phase, only the individuals of the same agent are chosen to crossover. Additionally, this operation is performed only on the transited branches. Similarly, the mutation operation is performed only on the branches transited during execution.

*Algorithm 1 Pseudo code of proposed method.*

1. **Initialize** population *Pop* consisting of the connections in graph structures.
2. **Group** the individuals in n groups. n computed by dividing population size by the number of agents.
3. While the termination condition is not fulfilled
    3.1. **for** each group
        3.1.1. Each agent performs one step according to its corresponding individual and the transited branches are obtained.
        3.1.2. If the agents have not reached the last step or the goal, go to 3.1.1.
        3.1.3. Calculate the fitness of the group (i.e., individual).
    3.2. Preserve the elite individuals $Pop_e$
4. **for** (i=1:(*populationSize*/2 – *numberOfEliteIndividuals*))
    4.1. Select two parents *P1* and *P2* from *Pop*.
    4.2. **for** each branch $b \in TB_{p1} \cup TB_{p2}$ **do**
        4.2.1. **if** (random_seed < $p_c$)
            4.2.1.1. Swap the connection of *b* between the parents corresponding to each agent.
5. **for** (i=1:*populationSize* – *numberOfEliteIndividuals*)
    5.1. **for** each branch $b \in TB_p$ **do**
        5.1.1. **if** (random_seed < $p_m$)
            5.1.1.1. Randomly change the connection of *b*.

## 4. Experimental results

To evaluate the performance of the proposed algorithm, we tested it using the Tileworld benchmark [34]. We compared this algorithm to GNP_simplified and SBGNP regarding average fitness, standard deviation, number of successful solutions, and p-value. This benchmark is widely used in GNP and other decision-making tasks for agent control problems [31, 33, 35-37]. Tileworld is a 2D grid-based environment consisting of agents, tiles, holes, and obstacles. The goal is to drop all the tiles into the holes in the shortest time possible. Controlling the agents requires them to cooperate effectively. The agents can sense their surrounding environment using judgment nodes and take actions based on the output of these judgment nodes via processing nodes. This research used seven judgment functions and four processing functions, as illustrated in Table 1. We can develop a program to control the agents by using these functions appropriately.

*Table 1. List of judgment and processing nodes utilized in the Tileworld environment.*

| Node type | Symbol | Description | outputs |
|---|---|---|---|
| Judgment nodes | WEF | What Exist Front | Agent, Tile, Hole, Obstacle, Nothing |
| | WER | What Exist Right | |
| | WEL | What Exist Left | |
| | WEB | What Exist Back | |
| | NTD | Nearest Tile Direction | Forward, Backward, Right, Left |
| | SNTD | Second Nearest Tile Direction | |
| | NHD | Nearest Hole Direction | |
| Processing nodes | MF | Move Forward | --- |
| | TR | Turn Right | |
| | TL | Turn Left | |
| | ST | STay | |

To illustrate the function of these operations, the results of their application under the conditions shown in Figure 4(a) are provided in Figure 4(b). This example demonstrates how forward, backward, left, and right directions are considered.

To evaluate the agents' performance, three key factors are taken into account: within a restricted number of steps, the agents should:

1) push as many tiles into holes as they can

2) achieve this using the fewest possible steps

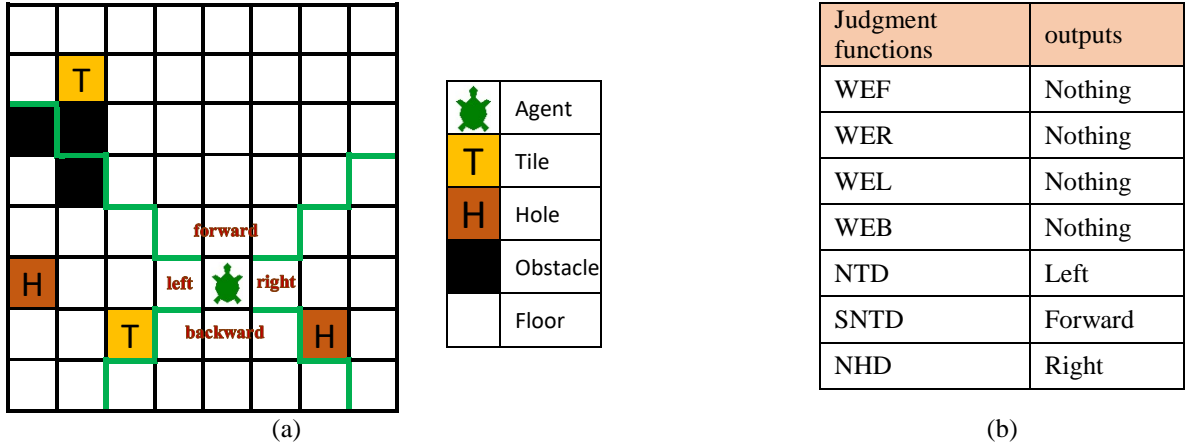

Figure 4. An illustrative example of judgment function outputs.

3) if any tiles are left, they should be positioned as close as possible to the nearest holes.

Therefore, the fitness function for the solutions in the Tileworld system is defined as follows:

$$f = w_1 \times DT + w_2 \times (IS - TS) + w_3 \times \left[\sum_{t=1}^{T}(ID_{(t)} - FD_{(t)})\right]$$

*DT* denotes the total number of tiles the agents have dropped into the holes. *IS* refers to the maximum number of steps allowed for an agent, as defined by the user. *TS* indicates the number of steps taken, while *T* represents the total number of tiles in the environment. $ID_{(t)}$ indicates the initial distance from tile *t* to its nearest hole, and $FD_{(t)}$ denotes the distance after *TS* steps. The parameters $w_1$, $w_2$, and $w_3$ are used to adjust the weights of the first, second, and third factors, respectively. In our experiment, we set these parameters to $w_1 = 100$, $w_2 = 1$, and $w_3 = 20$. Table 2 shows the hyperparameters used in the Tileworld benchmark. These are the same across all four tested algorithms.

*Table 2. Hyperparameters of the Tileworld benchmark*

| Parameter name | Parameter value |
| --- | --- |
| Population size | 100 |
| Elite size | 2 |
| Probability of crossover ($P_c$) | 0.4 |
| Probability of mutation ($P_m$) | 0.01 |
| Program size | 3 |
| Initial Steps | 60 |
| number of iterations | 1000 |

To conduct a more thorough analysis, we compared the performance of the algorithms on two distinct Tileworld environments, illustrated in Figure 5. The results in Figure 6 demonstrate the performance of the proposed method compared to GNP, SBGNP, and Simplified_GNP across two environments. In environment 1, the proposed method performs better than the other algorithms, showing a more rapid increase in fitness and achieving the highest mean fitness by the end of the evaluations. SBGNP achieves the second-highest results, with GNP and Simplified_GNP showing lower performance. These results indicate that the proposed method effectively constrains the search space and optimizes agent behavior more efficiently in this environment. In environment 2, the proposed method performs best again but with a relatively minor improvement over the other algorithms. The fitness curves of all four methods converge more closely, with GNP, SBGNP, and Simplified_GNP showing similar performance to the proposed method, particularly after around 40,000 evaluations. This suggests that while the proposed method achieves the best results in both environments, its advantage is more evident in the more complex scenario (environment 1). In contrast, the improvement is less significant in the simpler scenario (environment 2).

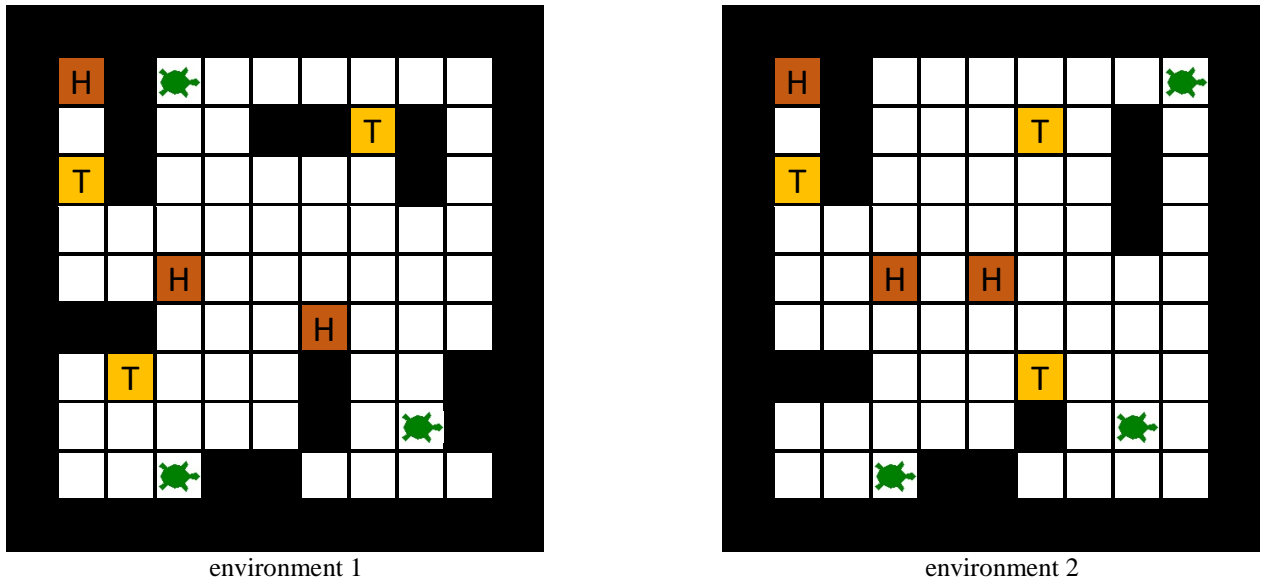

environment 1                                     environment 2

Figure 5. Tileworld environments used in the experiment

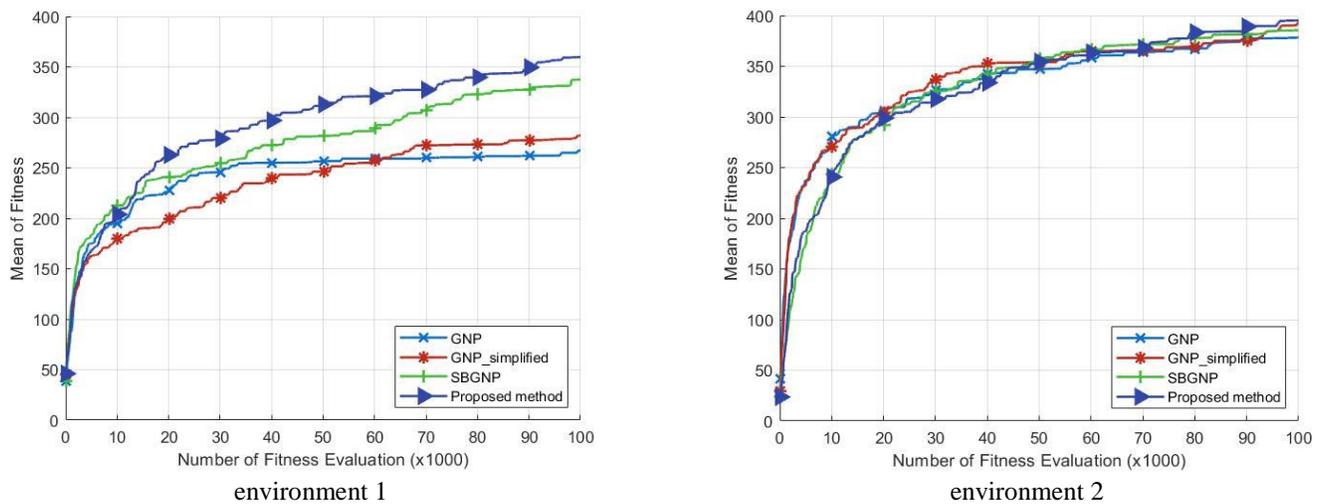

environment 1                                     environment 2

Figure 6. Fitness curve of the best solutions in two different environments

The statistical results presented in Table 3 and Table 4 compare the performance of various algorithms in two environments. In environment 1, the proposed method ranks first with an average fitness of 369.22, significantly outperforming GNP, GNP_simplified, and SB-GNP. The proposed method also achieves the highest number of successful solutions (13) and has a relatively low standard deviation of 78.52, indicating consistent performance. The p-value below 0.05 demonstrates statistical significance.

In environment 2, the proposed method again ranks first, achieving an average fitness of 395.22, slightly surpassing GNP_simplified, which ranks second with an average fitness of 393.04 and the highest number of successful solutions (26). The p-values in this environment indicate that while the proposed method and GNP_simplified exhibit better performance, they do not reach the same statistical significance level as in environment 1. Overall, the results demonstrate that the proposed method consistently outperforms the other algorithms across both environments, particularly in terms of average fitness and relatively in terms of the number of successful solutions, reinforcing its effectiveness in enhancing multiagent GNP performance.

*Table 3. Statistical results of algorithms in environment 1*

| Algorithm | Rank | Average | Standard Deviation | Number of Successful Solutions | p-value |
|---|---|---|---|---|---|
| GNP | 4 | 267.22 | 80.34 | 2 | 1.301E-07 |
| GNP_simplified | 3 | 282.16 | 93.60 | 4 | 4.895E-05 |
| SB-GNP | 2 | 337.44 | 84.05 | 9 | 3.296E-02 |
| **Proposed Method** | **1** | **369.22** | **78.52** | **13** | |

*Table 4. Statistical results of algorithms in environment 2*

| Algorithm | Rank | Average | Standard Deviation | Number of Successful Solutions | p-value |
|---|---|---|---|---|---|
| GNP | 4 | 378.60 | 101.59 | 23 | 0.1943 |
| GNP_simplified | 2 | 393.04 | 95.08 | 26 | 0.4516 |
| SB-GNP | 3 | 385.46 | 79.88 | 22 | 0.2591 |
| **Proposed Method** | **1** | **395.22** | **81.85** | **25** | |

## 5. Conclusion

GNP is an evolutionary algorithm that extends the tree structure of GP into a graph structure. During the evolutionary process, the connections between nodes change. This results in a large search space, making it difficult to find an optimal solution within a limited number of iterations, particularly when using a separate graph structure for each agent. In this study, inspired by GNP_simplified, we applied simplified operators to reduce the search space, leading to enhanced convergence and more efficient evolution in multiagent GNPs. By testing this algorithm on the Tileworld benchmark, we observed improvements in the fitness curve and the number of successful solutions. In future works, other improvements for multiagent environments can be explored. More intelligent approaches to reducing the search space could further improve the efficiency of the evolutionary process.